%% file: 2022MMSP_heimann.tex
\documentclass[conference, 9pt]{IEEEtran}
\IEEEoverridecommandlockouts
\usepackage{cite}
\usepackage{amsmath,amssymb,amsfonts}
\usepackage{algorithmic}
\usepackage{graphicx}
\usepackage{textcomp}
\usepackage{xcolor}
\usepackage{pgfplots}
\usepackage{tikz}  
\usepackage{tikz-3dplot} 
\usetikzlibrary{shapes,arrows, positioning, automata, shadows,arrows.meta,backgrounds,fit, calc}
\usepackage{tkz-fct, tikz-dimline}
\usepackage{subcaption}
\pgfplotsset{compat=1.17} 

\usepackage{multirow}
\def\BibTeX{{\rm B\kern-.05em{\sc i\kern-.025em b}\kern-.08em
    T\kern-.1667em\lower.7ex\hbox{E}\kern-.125emX}}
    
\newcommand*\circled[1]{\tikz[baseline=(char.base)]{
            \node[shape=circle,draw,inner sep=1pt] (char) {#1};}}
        
\usepackage{hyperref}

\newcommand\copyrighttext{%
	\footnotesize \textcopyright 2022 IEEE. Personal use of this material is permitted.
	Permission from IEEE must be obtained for all other uses, in any current or future
	media, including reprinting/republishing this material for advertising or promotional
	purposes, creating new collective works, for resale or redistribution to servers or
	lists, or reuse of any copyrighted component of this work in other works.
}
\newcommand\copyrightnotice{%
	\begin{tikzpicture}[remember picture,overlay]
		\node[anchor=south,yshift=10pt] at (current page.south) {\fbox{\parbox{\dimexpr\textwidth-\fboxsep-\fboxrule\relax}{\copyrighttext}}};
	\end{tikzpicture}%
}

\begin{document}

\title{Jointly Resampling and Reconstructing Corrupted Images for~Image Classification using Frequency-Selective~Mesh-to-Grid~Resampling}
\author{\IEEEauthorblockN{Viktoria Heimann, Andreas Spruck, and Andr\'e Kaup}
\IEEEauthorblockA{Multimedia Communications and Signal Processing\\
Friedrich-Alexander-Universität Erlangen-Nürnberg\\
Cauerstraße 7, 91058 Erlangen\\
Email: \{viktoria.heimann, andreas.spruck, andre.kaup\}@fau.de}\\
}

\maketitle
\copyrightnotice
\begin{abstract}
Neural networks became the standard technique for image classification throughout the last years. They are extracting image features from a large number of images in a training phase. In a following test phase, the network is applied to the problem it was trained for and its performance is measured. In this paper, we focus on image classification. The amount of visual data that is interpreted by neural networks grows with the increasing usage of neural networks. Mostly, the visual data is transmitted from the application side to a central server where the interpretation is conducted. If the transmission is disturbed, losses occur in the transmitted images. These losses have to be reconstructed using postprocessing. In this paper, we incorporate the widely applied bilinear and bicubic interpolation and the high-quality reconstruction Frequency-Selective Reconstruction (FSR) for the reconstruction of corrupted images. However, we propose to use Frequency-Selective Mesh-to-Grid Resampling (FSMR) for the joint reconstruction and resizing of corrupted images. The performance in terms of classification accuracy of EfficientNetB0, DenseNet121, DenseNet201, ResNet50 and ResNet152 is examined. Results show that the reconstruction with FSMR leads to the highest classification accuracy for most networks. Average improvements of up to 6.7~percentage points are possible for DenseNet121.  

\end{abstract}

%

\section{Introduction}
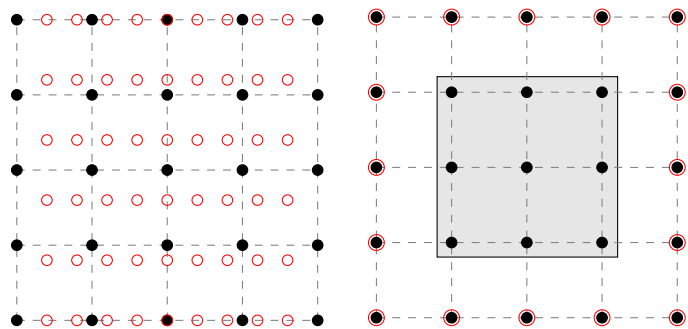
\begin{figure}[t]
\centering
\begin{subfigure}[t]{.45\columnwidth}
\input{figures/DatabaseNew}
\caption{\label{fig:resizemesh} Original points after resizing on mesh points. Resampling onto pixel grid necessary.}
\end{subfigure}
\qquad
\begin{subfigure}[t]{.45\columnwidth}
\input{figures/grid_with_hole}
\caption{\label{fig:patternloss} \textit{BLOCK} loss covers the center (gray). No original points in this area. Pixel grid must be reconstructed.}
\end{subfigure}
\caption{\label{fig:mesh}Original data given as red circles. The full pixel grid as goal of resizing and reconstruction given as black dots.}
\vspace{-.5cm}
\end{figure}

Neural networks are widely applied during the last years for image processing. They are especially used for interpreting images. A task that is easy for humans but difficult for machines. Neural networks are able to extract features from image content. They do not require the image content to be described analytically. They learn to extract the important image features by training. In the training phase, a large number of images is presented to the network. Thus, if only a small number of images is available, techniques such as data augmentation have to be applied in order to create a larger number of images for training. Data augmentation refers to the task of artificially creating additional data with new characteristics from the already available data. Typical techniques for data augmentation are geometric transformations such as zooming and rotating input images. Furthermore, most of the networks demand a fixed input size and thus, the input images have to be resized. Resizing on pixel level means, that pixels are transformed to a new position. Thereby, the resized pixels are mostly located on mesh positions that are not fixed to an integer grid. An example on pixel level is shown in Fig.~\ref{fig:resizemesh}. The red circles denote the resized pixels on mesh positions, whereas the black dots denote the pixel grid. The dashed lines support the presentation of the pixel grid. For a further processing, the resized pixels at mesh positions have to be resampled to the pixel grid.\\
Throughout the last years, efficient network architectures were developed \cite{imagenet, Tan_2019, Huang_2017, He_2016}. They provide good results for image classification. With the increasing usage of neural networks and the production of a growing number of images and videos, an increasing amount of visual data is interpreted by machines. Typically, the interpretation is conducted after transmitting the data to a central server in both, cloud and edge computing scenarios \cite{Cao_2021, Hassan_2019}. For transmission, the visual content has to be compressed. Current compression schemes typically separate the images into smaller blocks \cite{Sullivan_2012_HEVC, Bross_2021_VVC}. If the transmission is disturbed, only a subset of blocks can be transmitted properly. The remaining image parts have to be filled using reconstruction algorithms. The scenario is demonstrated on pixel level in Figure~\ref{fig:patternloss}. As in Figure~\ref{fig:resizemesh}, the black dots denote the pixel grid. The gray colored area in the middle represents a \textit{BLOCK} loss. The original data is given as red circles outside the gray area. The information for the black dots in the gray backed area has to be reconstructed from the red data points. If such an image should be analyzed by a neural network, the loss in the image has to be reconstructed first. Thereafter, the image is resized to the expected input size of the network. Reconstructing missing parts of images was mainly investigated for humans as observers and thus, the algorithms are optimized mainly in terms of Peak-Signal-to-Noise-Ratio (PSNR) and Structural Similarity Index Measurement (SSIM) \cite{Seiler_2015, Seiler_2010_ComplexFSE}. A high-quality reconstruction technique is Frequency-Selective Reconstruction (FSR) \cite{Seiler_2015}. For the resizing step, a classical interpolation such as bilinear or bicubic interpolation can typically be incorporated. However, this preprocessing pipeline consists of two processing parts that might both introduce errors to the final image. Hence, we propose to resize the image including its losses and combine reconstruction and resampling into one computation step. Thus, we aim at minimizing the error that was made in the preprocessing of a neural network and thereby, increase the classification accuracy. We incorporate Frequency-Selective Mesh-to-Grid Resampling (FSMR) for this purpose. FSMR could already proof to be beneficial as a preprocessing method for the resizing of test images \cite{Spruck_2022}. \\
We will briefly introduce FSMR in the upcoming section. Thereafter, we present our proposed procedure in Sec.~\ref{sec:proposed}. It follows our experimental setup in Sec.~\ref{sec:setup}. Next, we show the results of our extensive experiments in Sec.~\ref{sec:exp}. Finally, the paper closes with a conclusion.  

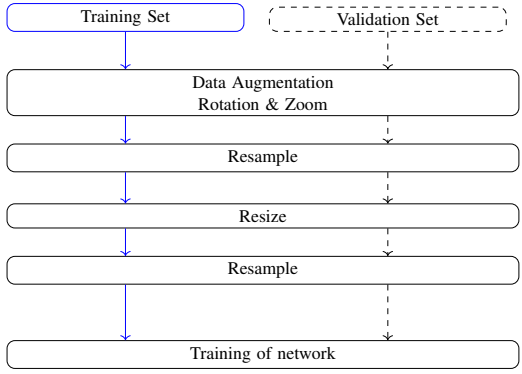
\begin{figure}[t]
\centering
\resizebox{0.85\columnwidth}{!}{
	\centering
\input{figures/flow_graph}		 
}
\caption{\label{fig:trainingsetup}Training set up as flow chart.}
\end{figure}

\section{\label{sec:fsmr} Frequency-Selective Mesh-to-Grid Resampling}
A common approach in image signal processing is to view the image in the spectral domain. Images can be represented in the spatial domain as well as in the spectral domain. The spectral domain of images can be reached by transforming the image using Discrete Fourier or Discrete Cosine Transform (DCT). This representation gives the frequency distribution of the image. Natural images can usually be represented in terms of only few basis functions. If pixels are located on mesh positions due to, e.g., a geometrical transform, the spectral discrete transforms cannot be carried out directly any more. Thus, the frequencies contained in the geometrically transformed image have to be estimated with a model. The main assumption of our model is that an image \(f[m,n]\) can be represented in terms of a weighted superposition of basis functions \(\varphi_{k, l}[m, n]\)
\begin{equation}
\label{eq:image}
f[m, n] = \sum_{k, l \in \mathcal{K}} c_{k, l} \varphi_{k, l}[m, n],
\end{equation}
where the indexes \(k\) and \(l\) denote the frequency index in horizontal and vertical direction, respectively. They are chosen from the set of all possible basis functions \(\mathcal{K}\) at every coordinate position \([m,n]\). The model itself is set to zero in the beginning, i.e., \(g^{(0)}[m,n]\equiv 0\). During the model generation process the best fitting basis function \(\varphi_{u, v}[m, n]\) with its according expansion coefficient \(\hat{c}_{u, v}\) is added in the current iteration \(\nu\) to the model from the previous iterations \(g^{(\nu -1)}[m, n]\)
\begin{equation}
\label{Eq:modelGeneration}
g^{(\nu)}[m, n] = g^{(\nu -1)}[m, n] + \hat{c}_{u, v} \varphi_{u, v}[m, n]. 
\end{equation}
As the aim of the generated model is to estimate the original image signal as good as possible, the differences between the model and the original have to be minimized. Thus, the residual \(r^{(\nu)}\) is given as
\begin{equation}
\label{Eq:Residual}
r^{(\nu)} = f[m,n] - g^{(\nu)} [m,n].
\end{equation}
The minimization of the residual is then formulated in terms of the residual energy
\begin{equation}
E^{(\nu)} = \sum_{(m,n)} w[m,n]\left(r^{(\nu)}[m,n]\right)^2,
\end{equation}
where a spatial weighting function \(w[m,n]\) is incorporated additionally. It is defined as an isotropically decaying window function with its center in the middle of the currently processed block. This enables the model to prefer center regions and adapt to local characteristics of the image. The indexes \({(u, v)}\) of the best fitting basis function that is added in the current iteration can be determined using the residual energy according to 
\begin{equation}
\label{eq:bestfitting}
{(u, v)} = \underset{{(k, l)}}{\mathrm{argmax}} \left( \Delta E_{k, l}^{(\nu)} w_{f}[k,l] \right).
\end{equation}

\begin{figure}[tp]
\begin{subfigure}[t]{.47\columnwidth}
\centering
	\resizebox{\columnwidth}{!}{
\input{figures/test_setup_old}		 
}
\caption{\label{fig:testsetup1} Common sequential {approach}.}
\end{subfigure}
\begin{subfigure}[t]{.45\columnwidth}
\centering
	\resizebox{\columnwidth}{!}{
\input{figures/test_setup_proposed}		 
}
\caption{\label{fig:testsetup2} Proposed joint approach.}
\end{subfigure}
\caption{Testing set ups as flow chart.}
\vspace{-.6cm}
\end{figure}
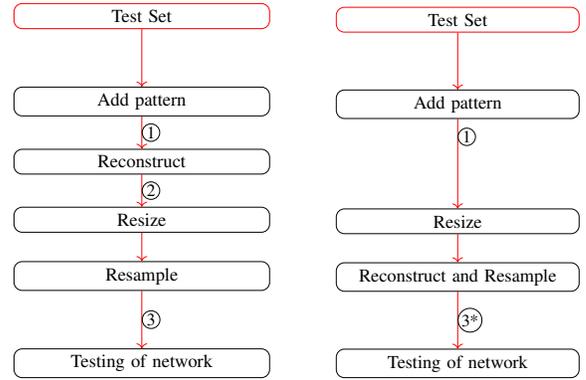

In \eqref{eq:bestfitting}, a spectral weighting function \(w_{f}[k,l]\) is incorporated. It favors low frequencies as it could be shown that natural images are mainly composed of low frequency components~\cite{2000_Lam_DCTCoeffAnalysis}. High frequencies tend to produce artifacts such as ringing. Nevertheless, if high frequencies are dominant in the image, they can still be included in the model. In every iteration, the basis function is chosen that maximizes the residual energy reduction the most as the gap between original signal and model should be closed as fast as possible. \\
Finally, the model is evaluated for the goal coordinates \([o,p]\) on the grid 
\begin{equation}
\label{Eq:evaluationonmesh}
f[o,p] = \sum_{k, l \in \mathcal{K}} \hat{c}_{k, l} \varphi_{k, l}[o, p].
\end{equation}  
For the final evaluation the estimated expansion coefficient of every possible basis function is multiplied with the according basis function and summed up. \\
FSMR was shown to be a high performing method for various resampling scenarios such as affine transforms \cite{Heimann_2020_MMSP} and motion compensated frame-rate up-conversion \cite{Heimann_2021}.

\section{Proposed Approach}
\label{sec:proposed}
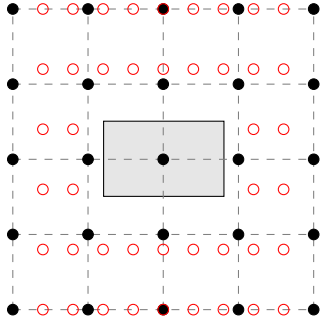
\begin{figure}[t]
\centering
\input{figures/grid_with_hole_resized}
\caption{\label{fig:patternlossresized} Original points after resizing on mesh points (red circles). Resized \textit{BLOCK} loss covers the center (gray rectangle). No original points in this loss area. Pixel grid must be resampled and reconstructed (black dots).}
\vspace{-.7cm}
\end{figure}

In this study, we assume a transmission scenario where losses occur. The transmitted images should be classified into object categories by a neural network. The network itself was trained for non-corrupted images. Thus, the losses have to be reconstructed and the images have to be resized to the desired input size of the neural network. Common approaches handle this task sequentially. The losses are first reconstructed and subsequently the images are resized. The sequential approach is depicted in Fig.~\ref{fig:testsetup1}. The test set represents the images that should be transmitted. Thereafter, a pattern is added that simulates the losses during the transmission. A corrupted image is generated that serves as the received image in our scenario. For this corrupted image, the losses are reconstructed first. An example of a corrupted image that has to be reconstructed is given on pixel level in Fig.~\ref{fig:patternloss}. Secondly, the image is resized and the mesh pixels are resampled onto the desired grid points. The resampling is depicted in Fig.~\ref{fig:resizemesh}. The sequential approach requires two interpolation steps, reconstruction and resampling. Thus, interpolation artifacts might be induced twice in the reconstructed and resized image signal. \\
Hence, we propose to jointly reconstruct the loss and resize the image to the desired input size of the neural network. Our proposed processing pipeline is depicted in Fig.~\ref{fig:testsetup2}. As for the common sequential approach, a loss pattern is added to each image of the test set yielding the corrupted received image in the assumed transmission scenario. The corrupted image is directly resized to the desired input image size of the neural network. The thereby resulting image is depicted on pixel level in Fig.~\ref{fig:patternlossresized}. The image shows mesh points in red that have to be resampled to the black dots on grid positions. The loss area is given in gray. The figure results from resizing the image depicted in Fig.~\ref{fig:patternloss} to the desired input image size of the network. The resizing does not necessarily keep the aspect ratio of the original image. Thus, the figure can be interpreted as the combination of Fig.~\ref{fig:resizemesh} and Fig.~\ref{fig:patternloss}. In the final resampling step in Fig.~\ref{fig:testsetup2}, the mesh points are resampled to the grid points in Fig~\ref{fig:patternlossresized}. Thereby, the loss is reconstructed and the image is resized in one joint step. As only one interpolation step is necessary in our proposed scenario, it can be expected that less interpolation artifacts occur. Furthermore, a higher quality can be anticipated as there are more original image points available for the reconstruction of the center point than it was the case in Fig.~\ref{fig:patternloss}.

\section{Experimental Setup}
\label{sec:setup}

\begin{figure*}[t!]
\begin{subfigure}[t]{.33\columnwidth}
\includegraphics[width=\columnwidth]{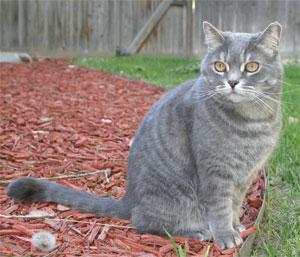}
\caption{\label{fig:orig} Original.}
\end{subfigure}
\begin{subfigure}[t]{.33\columnwidth}
\includegraphics[width=\columnwidth]{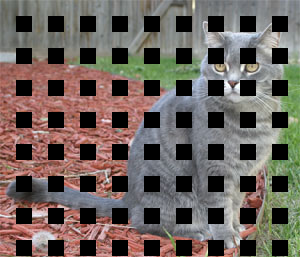}
\caption{\label{fig:orig_loss} \textit{BLOCK} loss \circled{\small1}.}
\end{subfigure}
\begin{subfigure}[t]{.33\columnwidth}
\includegraphics[width=\columnwidth]{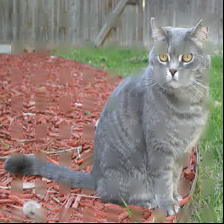}
\caption{\label{fig:fse} FSR \circled{\small3}.}
\end{subfigure}
\begin{subfigure}[t]{.33\columnwidth}
\includegraphics[width=\columnwidth]{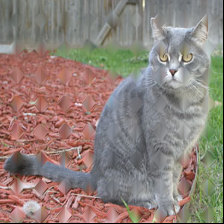}
\caption{\label{fig:cubo} cub \circled{\small3}.}
\end{subfigure}
\begin{subfigure}[t]{.33\columnwidth}
\includegraphics[width=\columnwidth]{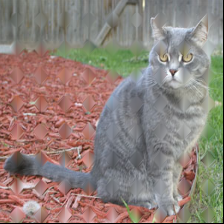}
\caption{\label{fig:lino}lin \circled{\small3}.}
\end{subfigure}
\begin{subfigure}[t]{.33\columnwidth}
\includegraphics[width=\columnwidth]{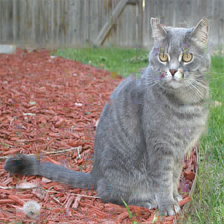}
\caption{\label{fig:fsmr}FSMR \circled{\small3*}.}
\end{subfigure}
\caption{\label{fig:inpainting} Development of the \textit{British Shorthair 239} image throughout the test path. Original image and image with loss are of size \(300\times257\)~pixels, all remaining images are of size \(224\times224\)~pixels. Circled numbers denote the position in Figure~\ref{fig:testsetup1} and~\ref{fig:testsetup2}. Artifacts from Fig.~\ref{fig:fse} onwoards best be viewed enlarged.}
\vspace{-.5cm}
\end{figure*}

The experimental setup has to be separated into two phases, the training phase and the testing phase of a network. We will first carry out the training phase and thereafter, present our proposed testing procedure in comparison to the state-of-the-art procedure. For this purpose, also the incorporated datasets are separated into three sets, training, validation and test set. The training set consists of the largest number of images. Training and validation set are used in the training phase whereas the test set is solely used in the testing phase. The data sets, neural networks and processing techniques remain the same for training and testing phase and thus, are explained in more detail in Sec.~\ref{sec:dset}, \ref{sec:recon}, and \ref{sec:nn}, respectively.

\subsection{Training Phase}
The training and validation sets of a dataset are incorporated during the training phase of a network. In Fig.~\ref{fig:trainingsetup}, the path of the training set is denoted in blue and the path of the validation set is given as dashed lines. For a better generalization of a neural network, the images are usually augmented. Both datasets are augmented using rotation and zoom operations and thus, the number of training images is tripled. As presented in \cite{Spruck_2022}, the augmented data has to be resampled in the following step in order to resample the mesh points onto the regularly spaced grid positions. After the augmentation step, the images have to be resized to the input size of the neural networks. All networks incorporated in this study require an input size of \(224\times224\)~pixels. As the resizing step also yields pixels on mesh positions, resampling has to be carried out for the second time. For both resampling steps, we used the same interpolation method. Finally, the network is trained using the prepared datasets. \\
\subsection{Testing Phase}
\label{sec:testphase}
For testing the network, the test set is incorporated. We assume the test set to be distorted by losses and thus, a reconstruction of the losses is necessary. Furthermore, the images have to be resized to the input size of the network. There are two ways of reconstructing and resizing images in this application. \\
The state-of-the-art preprocessing pipeline is shown in Fig.~\ref{fig:testsetup1}. In addition, Fig.~\ref{fig:inpainting} shows how an exemplary image evolves along the processing pipeline. At the beginning, an arbitrary image such as in Fig.~\ref{fig:orig} is taken. This image is distorted by a loss pattern as given in Fig.~\ref{fig:orig_loss}. Next, the losses are reconstructed using a reconstruction method such as FSR. The image still holds the original size which is not the expected input size of the network of \(224\times224\)~pixels. Thus, the reconstructed image is resized accordingly. Following the common sequential approach as depicted in Fig.~\ref{fig:testsetup1}, examples for reconstructed and resized images are depicted in Figs.~\ref{fig:fse}-\ref{fig:lino}. \\
Our proposed pipeline is demonstrated in Fig.~\ref{fig:testsetup2}. Once again, a pattern is added to the test image. However in our joint approach, the loss is not reconstructed directly. We apply the resizing step on the image with losses. Thus, the resizing step shifts the available pixels onto mesh positions. Hence, the reconstruction and resampling is conducted in one joint step. This holds the advantage that only one processing step is required in the preprocessing of the test set before the network is evaluated. The final result for our proposed joint approach incorporating FSMR is shown in Fig.~\ref{fig:fsmr}. \\
The used disortion patterns are described and shown in the upcoming section.

\subsection{Patterns}
\begin{figure}[t!]
\begin{subfigure}[t]{.32\columnwidth}
\centering
\includegraphics[width=\columnwidth]{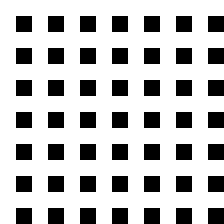}
\caption{\label{fig:block}\textit{BLOCK}}
\end{subfigure}
\begin{subfigure}[t]{.32\columnwidth}
\centering
\includegraphics[width=\columnwidth]{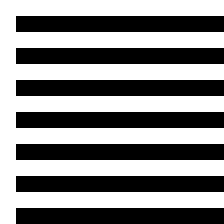}
\caption{\label{fig:line} \textit{LINE}}
\end{subfigure}
\begin{subfigure}[t]{.32\columnwidth}
\centering
\includegraphics[width=\columnwidth]{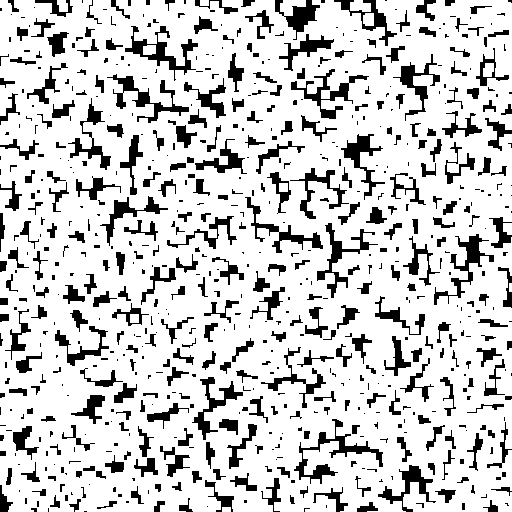}
\caption{\label{fig:rand}\textit{RAND}}
\end{subfigure}
\caption{\label{fig:pattern} Loss patterns of size \(224\times224\)~pixels.}
\vspace{-.7cm}
\end{figure}

We incorporate three different patterns for losses. There is the \textit{BLOCK} pattern first. It shows square losses which are regularly distributed over the whole image. An excerpt of \(224\times224\) pixels of the \textit{BLOCK} pattern is given in Figure~\ref{fig:block}. Black parts in the figure denote the covered areas in the image. Furthermore, there is the \textit{LINE} pattern. It is given in Figure~\ref{fig:line}. Horizontal lines are masking the image in regular distances. As a third category, we incorporate a random pattern given as \textit{RAND} in Figure~\ref{fig:rand}. Additionally, we created the category \textit{ALL}. This category consists of all loss patterns, \textit{BLOCK}, \textit{LINE}, and \textit{RAND}. 

\subsection{Datasets}
\label{sec:dset}
For our studies, we incorporated two widely used datasets, namely the \textit{Caltech101} \cite{caltech101} and the \textit{Oxford-IIIT-Pets} \cite{oxford_3t_pet}. As mentioned above, we split the two datasets in test, validation and training set first. For this separation, we chose an 80 -- 10 -- 10 distribution meaning that 80\% of the dataset is used for training, 10\% of each class are used for validation during training and the remaining 10\% are used for evaluation after the training process. \\
The \textit{Caltech101} set is a dataset of natural images that contains 102 classes. The classes show a large variety of everyday objects. They range from animals over food to different means of transport. Most of the classes hold around 50 images. Nevertheless, there are classes that hold just 40 or up to 800 images. In order to establish a fair comparison and to cope with the unbalanced dataset, we used a uniformly distributed test set. We used approximately 10\% of every class for testing. Thus, 5 samples from each class are incorporated in the test set which helps to keep the test results comparable. \\
Our second incorporated dataset, the \textit{Oxford-IIIT-Pets} shows natural images again. It consists of 12 breeds of cats and of 25 breeds of dogs. Thus, in total it contains 37 different classes of pets. The \textit{Oxford-IIIT-Pets} dataset contains approximately 200 images per class, although one class just contains 191 images. Following our dataset split scheme from above, 20 images of every class are used for testing. The images within one class vary clearly in terms of perspective, aspect ratio, illumination and background. On the other side, the breeds seem to be difficult to be separated sometimes. Thus, the differences between the classes might be small and thereby making this dataset a challenging one.

\subsection{Reconstruction and Resampling}
\label{sec:recon}
We trained five neural networks from scratch. The five networks are presented in more detail in the upcoming section. For the training procedure, we applied data augmentation techniques. We rotated and zoomed the input images. After the data augmentation, we resized all images to the desired input size of \(224\times224\) pixels. Rotation, zoom and resizing demand for an interpolation of pixels that are not located on the pixel grid but rather on mesh positions. These mesh points have to be resampled to the pixels on the grid using interpolation. For the training dataset, we used bilinear, and bicubic interpolation and the presented FSMR method for the resampling step. For both resampling steps in the training and validation path, the same interpolation technique was incorporated. The networks are trained for each interpolation technique separately. Thus, also the interpolation technique for training and validation was the same. \\
During the test phase of the networks, we did not apply data augmentation methods. We rather assumed the images to show transmission losses that have to be reconstructed. In the state-of-the-art approach given in Fig.~\ref{fig:testsetup1}, we incorporated three techniques for reconstruction. There is the high-quality Frequency-Selective Reconstruction \cite{Seiler_2015} first. Although, bilinear and bicubic interpolation are usually not preferred for the reconstruction of large block patterns for human users, we incorporated them here for two reasons. First, research has shown that neural networks can profit from a smooth image interpolation such as bilinear interpolation \cite{Spruck_2022}. Second, bilinear and bicubic interpolation are implemented in typical machine learning platforms such as TensorFlow \cite{tensorflow} and PyTorch \cite{pytorch} and thus, are often incorporated as preprocessing methods for input data of neural networks. We refer to the use of bilinear interpolation for reconstruction and resampling for the state-of-the-art test setup as \textit{lin}. An example is given in Fig.~\ref{fig:lino}. The reconstructed blocks are still clearly visible. If for both operations, bicubic interpolation is incorporated, we denote it as \textit{cub} which results in an image as in Fig.~\ref{fig:cubo}. Also in this case, the structure of the loss pattern is easily observable. If FSR is used for the reconstruction and bicubic interpolation is used for resampling, it is given as FSR and shown in Fig.~\ref{fig:fse}. Although FSR is a high-quality reconstruction technique, the block pattern is still slightly visible in the image.  FSMR is used as resampling technique for our proposed test setup as explained in Sec.~\ref{sec:proposed} and given in Fig.~\ref{fig:testsetup2}. An example is shown in Fig.~\ref{fig:fsmr}. The resizing and reconstruction quality appears to be much better as the block structure of the loss pattern is not visible any more. Only some minor artifacts occur. 
\vspace{-.2cm}
\subsection{Neural Networks}
\label{sec:nn}
For our experiments, we incorporated five commonly used image classification networks namely EfficientNetB0 \cite{Tan_2019}, DenseNet121 and DenseNet201 \cite{Huang_2017}, as well as ResNet50 and ResNet152 \cite{He_2016}. They all showed very good results for image classification tasks. Furthermore, it was important for us to demonstrate the influence of the reconstruction method onto networks with a varying number of weights. The class of EfficientNets \cite{Tan_2019} is easily scalable. In our studies, we incorporate the smallest network from the class of EfficientNets, i.e., EfficientNetB0. It holds 5.3 million parameters. In addition, we incorporate DenseNet121 and DenseNet201 \cite{Huang_2017}. The class of DenseNets incorporates a dense connection from one layer to every other layer in feed forward direction. Thus, the feature maps from all preceding layers are available to all upcoming layers. Our incorporated DenseNet121 holds 8 million parameters. DenseNet201 is deeper and thus, holds 20 million parameters. The third class of incorporated neural networks into our studies is the class of ResNets. We used ResNet50 and ResNet152 \cite{He_2016}. ResNets are known for deep networks that optimize the residuals during the training phase. ResNet50 holds 25 million parameters, whereas ResNet152 is the largest network in our study as it holds 60 million parameters.

\vspace{-.2cm}
\section{Evaluation}
\label{sec:exp}

\begin{table*}[tp]
	\centering
	\caption{Classification accuracy in \% for \textit{Caltech101}. Trained without pattern. Only test with pattern loss. Best per pattern, train interpolation and network is given in bold. Best per pattern and network is additionally underlined.}
	\label{tab:results_caltech}
	\resizebox{\textwidth}{!}{
		\begin{tabular}{|c|c|cccc|cccc|cccc|cccc|cccc|} \hline
			  &									& \multicolumn{4}{c}{EfficientNetB0} 	& \multicolumn{4}{|c}{DenseNet121}		& \multicolumn{4}{|c}{DenseNet201} & \multicolumn{4}{|c}{ResNet50}		& \multicolumn{4}{|c|}{ResNet152} \\ \cline{3-22}
				& Train & \multicolumn{4}{c}{Test Interpolation} & \multicolumn{4}{|c}{Test Interpolation} & \multicolumn{4}{|c}{Test Interpolation} & \multicolumn{4}{|c}{Test Interpolation} & \multicolumn{4}{|c|}{Test Interpolation}  \\
 Pattern &	 Interp.	& 	lin 	& cub 	& FSR & FSMR 		& 	lin 	& cub 	& FSR & FSMR& 	lin 	& cub 	& FSR & FSMR& 	lin 	& cub 	& FSR & FSMR& 	lin 	& cub 	& FSR & FSMR \\ \hline
\multirow{3}{*}{\textit{BLOCK}} 
&  lin & 56.3 & 56.5 & \textbf{60.6} & 57.3 & 57.1 & 56.5 & 61.0 & \textbf{64.3} & 60.6 & 61.4 & 65.1 & \textbf{67.5} & 63.9 & 63.5 & 67.8 & \textbf{69.0} & 63.7 & 63.3 & 68.8 & \textbf{71.0} \\ \cline{2-22} 
& cub & 60.2 & 59.0 & \textbf{63.9} & 59.8 & 58.2 & 57.3 & 61.0 & \textbf{62.7} & 64.9 & 63.5 & 67.8 & \textbf{71.6} & 62.9 & 63.1 & \textbf{67.1} & 66.7 & 63.3 & 60.2 & 67.1 & \textbf{68.6} \\ \cline{2-22} 
& FSMR & 48.6 & 49.2 & 50.0 & \underline{\textbf{65.5}} & 57.3 & 57.3 & 59.0 & \underline{\textbf{67.5}} & 63.7 & 64.3 & 69.4 & \underline{\textbf{76.3}} & 59.2 & 59.8 & 63.5 & \underline{\textbf{70.0}} & 62.4 & 62.5 & 66.5 & \underline{\textbf{71.2}} \\
\hline	 \hline								
\multirow{3}{*}{\textit{LINE}}  & 
 lin  & 48.8 & 48.2 & \textbf{55.5} & 47.8 & 42.4 & 39.8 & 44.5 & \textbf{55.5} & 48.6 & 49.0 & 50.6 & \textbf{60.4} & 52.9 & 51.8 & 56.1 & \textbf{64.3} & 54.1 & 53.7 & 55.9 & \textbf{59.8} \\ \cline{2-22} 
& cub & 54.5 & 55.7 & \textbf{56.3} & 54.7 & 45.1 & 45.9 & 41.6 & \textbf{52.9} & 52.0 & 49.8 & 53.5 & \textbf{63.3} & 51.2 & 50.8 & 53.9 & \textbf{59.8} & 55.3 & 55.3 & 54.1 & \textbf{61.8} \\ \cline{2-22} 
& FSMR & 42.7 & 42.2 & 42.9 & \underline{\textbf{57.8}} & 47.6 & 47.6 & 42.9 & \underline{\textbf{62.0}} & 52.4 & 51.6 & 53.9 & \underline{\textbf{70.6}} & 48.6 & 47.8 & 49.0 & \underline{\textbf{64.7}} & 52.2 & 51.4 & 59.4 & \underline{\textbf{66.5}} \\
\hline \hline			 																	
\multirow{3}{*}{\textit{RAND}} & 
 lin  & 57.5 & 57.1 & \textbf{62.4} & 57.8 & 61.6 & 61.8 & 66.5 & \textbf{68.0} & 62.2 & 62.5 & \textbf{68.6} & 67.5 & 64.3 & 64.7 & \textbf{69.6} & \textbf{69.6} & 64.9 & 63.9 & 70.2 & \textbf{70.6} \\ \cline{2-22} 
 & cub & 59.0 & 58.8 & \underline{\textbf{63.9}} & 58.2 & 61.6 & 61.4 & \textbf{66.3} & 64.7 & 67.6 & 67.5 & 72.0 & \textbf{73.7} & 63.5 & 64.9 & 66.5 & \textbf{67.5} & 62.7 & 63.5 & 67.8 & \textbf{67.8} \\ \cline{2-22} 
 & FSMR & 50.2 & 51.2 & 54.9 & \textbf{62.4} & 59.4 & 59.6 & 63.1 & \underline{\textbf{68.2}} & 67.1 & 66.5 & 71.8 & \underline{\textbf{77.6}} & 61.8 & 61.2 & 65.1 & \underline{\textbf{72.0}} & 64.5 & 64.3 & 70.2 & \underline{\textbf{72.4}} \\
\hline \hline
\multirow{3}{*}{\textit{ALL}} & 
lin  & 54.0 & 54.3 & \textbf{59.2} & 54.5 & 53.7 & 52.7 & 57.3 & \textbf{62.6} & 57.4 & 57.8 & 61.2 & \textbf{65.0} & 60.6 & 60.2 & 64.5 & \textbf{67.3} & 61.4 & 60.5 & 65.2 & \textbf{67.5} \\  \cline{2-22} 
& cub & 57.9 & 58.0 & \textbf{62.0} & 57.5 & 55.4 & 55.2 & 56.6 & \textbf{60.0} & 61.6 & 60.5 & 64.6 & \textbf{69.3} & 58.9 & 59.5 & 61.8 & \textbf{64.4} & 60.3 & 60.2 & 63.0 & \textbf{65.8} \\  \cline{2-22} 
& FSMR & 47.0 & 47.6 & 49.5 & \underline{\textbf{62.2}} & 54.9 & 55.1 & 55.0 & \underline{\textbf{66.0}} & 61.7 & 61.6 & 65.7 & \underline{\textbf{74.6}} & 56.3 & 56.4 & 59.4 & \underline{\textbf{68.8}} & 59.6 & 59.3 & 65.7 & \underline{\textbf{70.2}} \\
\hline				
\hline
\hline
\multicolumn{2}{|r|}{Average} & 53.1 & 53.2 &56.8 & \textbf{58.0} & 54.5 & 54.2 & 56.2 & \textbf{62.9} & 60.0 & 59.7 & 63.7 & \textbf{69.8} & 58.7 & 59.2 & 62.0 & \textbf{67.0} & 60.4 & 59.8 & 64.5 & \textbf{67.8} \\
\hline										
		\end{tabular}
					}
\end{table*}

\begin{table*}[t]
	\centering
	\caption{Classification accuracy in \% for \textit{Oxford-IIIT-Pets}. Trained without pattern. Only test with pattern loss. Best per pattern, train interpolation and network is given in bold. Best per pattern and network is additionally underlined.}
	\label{tab:results_pet}
	\resizebox{\textwidth}{!}{
		\begin{tabular}{|c|c|cccc|cccc|cccc|cccc|cccc|} \hline
			  &									& \multicolumn{4}{c}{EfficientNetB0} 	& \multicolumn{4}{|c}{DenseNet121}		& \multicolumn{4}{|c}{DenseNet201} & \multicolumn{4}{|c}{ResNet50}		& \multicolumn{4}{|c|}{ResNet152} \\ \cline{3-22}
				& Train & \multicolumn{4}{c}{Test Interpolation} & \multicolumn{4}{|c}{Test Interpolation} & \multicolumn{4}{|c}{Test Interpolation} & \multicolumn{4}{|c}{Test Interpolation} & \multicolumn{4}{|c|}{Test Interpolation}  \\
 Pattern &	 Interp.	& 	lin 	& cub 	& FSR & FSMR 		& 	lin 	& cub 	& FSR & FSMR& 	lin 	& cub 	& FSR & FSMR& 	lin 	& cub 	& FSR & FSMR& 	lin 	& cub 	& FSR & FSMR \\ \hline
\multirow{3}{*}{\textit{BLOCK}} 
&  lin & \underline{\textbf{68.0}} & 66.8 & 66.6 & 62.2 & \textbf{84.2} & 83.6 & 82.0 & 83.6 & 90.3 & 89.9 & 88.9 & \textbf{90.4} & 81.1 & \textbf{81.8} & 81.1 & 81.5 & 81.8 & 82.3 & 81.4 & \textbf{82.8} \\ \cline{2-22} 
& cub & 65.8 & \textbf{66.9} & 66.4 & 63.1 & 91.1 & \underline{\textbf{91.4}} & 90.9 & 91.1 & 91.5 & 91.5 & \underline{\textbf{91.9}} & 91.5 & 83.4 & 83.4 & 82.6 & \underline{\textbf{83.6}}& 84.9 & \textbf{85.5} & 84.1 & 84.9 \\ \cline{2-22} 
& FSMR & 62.3 & 60.5 & 62.0 & \textbf{65.0} & 90.0 & 90.4 & 88.6 & \underline{\textbf{91.4}} & 87.2 & 88.1 & 88.0 & \textbf{88.8} & 75.5 & 76.1 & 73.9 & \textbf{76.6} & 86.4 & 86.9 & 86.6 & \underline{\textbf{87.3}} \\
\hline				\hline					
\multirow{3}{*}{\textit{LINE}}  & 
 lin & \underline{\textbf{63.2}} & 60.8 & 61.8 & 55.7 & \textbf{75.0} & 74.3 & 65.4 & 68.9 & 83.8 & \textbf{85.0} & 79.1 & 84.9 & 76.6 & \underline{\textbf{77.8}} & 70.0 & 74.1 & 76.5 & 77.0 & 71.1 & \textbf{77.8} \\ \cline{2-22} 
 & cub & 61.6 & 61.6 & 58.8 & \textbf{62.2} & 81.9 & \underline{\textbf{84.1}} & 76.6 & 80.0 & 85.5 & \underline{\textbf{86.5}} & 79.3 & 82.3 & 74.7 & \textbf{75.0} & 65.4 & 73.1 & 78.8 & \textbf{79.7} & 73.0 & 78.6 \\ \cline{2-22} 
 & FSMR & 58.1 & 59.6 & 54.5 & \textbf{61.4} & 81.6 & 81.8 & 74.7 & \textbf{82.8} & 79.9 & 80.9 & 74.6 & \textbf{82.7} & 65.8 & 67.6 & 58.4 & \textbf{68.4} & 80.9 & \underline{\textbf{81.8}} & 74.3 & 80.3 \\
\hline			 	\hline																
\multirow{3}{*}{\textit{RAND}} & 
lin & \underline{\textbf{68.5}} & 66.5 & 69.2 & 62.2 & \textbf{88.2} & 87.3 & 83.4 & 85.7 & 91.1 & \textbf{91.4} & 90.8 & 91.1 & 83.5 & \textbf{83.6} & 82.7 & \textbf{83.6} & 80.9 & 82.6 & \textbf{82.7} & 81.6 \\  \cline{2-22} 
& cub & 66.8 & 66.9 & \textbf{67.6} & 65.3 & 92.3 & 92.7 & 90.7 & \underline{\textbf{92.8}} & 91.9 & \underline{\textbf{92.4}} & 91.6 & 92.3 & 84.3 & 85.0 & 84.2 & \underline{\textbf{85.3}} & \textbf{86.4} & \textbf{86.4} & 85.5 & \textbf{86.4} \\ \cline{2-22} 
 & FSMR & 63.9 & 62.4 & 61.4 & \textbf{65.5} & 90.8 & \textbf{91.5} & 90.1 & \textbf{91.5} & 88.0 & 88.8 & 88.8 & \textbf{89.1} & 78.9 & \textbf{79.7} & 77.4 & 79.5 & 87.6 & 87.4 & 87.7 & \underline{\textbf{87.8}} \\
\hline \hline
\multirow{3}{*}{\textit{ALL}} & 
  lin & 53.8 & 52.2 & \underline{\textbf{65.9}} & 59.6 & 66.7 & 66.2 & 77.1 & \textbf{79.4} & 71.4 & 71.7 & 86.1 & \textbf{88.5} & 65.1 & 65.4 & 77.7 & \textbf{79.8} & 65.0 & 65.5 & 77.9 & \textbf{80.7} \\  \cline{2-22} 
 &cub & 52.4 & 52.7 & \textbf{64.1} & 63.3 & 71.5 & 72.4 & 86.1 & \textbf{87.7} & 72.3 & 72.6 & 87.9 & \underline{\textbf{89.0}} & 66.0 & 66.0 & 77.3 & \underline{\textbf{80.8}} & 68.0 & 68.2 & 80.8 & \underline{\textbf{83.1}} \\  \cline{2-22} 
 &FSMR & 50.0 & 49.3 & 59.4 & \textbf{63.9} & 70.8 & 71.1 & 84.6 & \underline{\textbf{88.4}} & 68.5 & 69.3 & 83.8 & \textbf{87.0} & 59.3 & 60.1 & 69.3 & \textbf{74.9} & 68.7 & 68.8 & 82.7 & \textbf{85.0} \\
\hline		
\hline
\hline
\multicolumn{2}{|r|}{Average} & 61.2 & 60.5 & \textbf{63.1} & 62.5 & 82.0 & 82.2 & 82.5 & \textbf{85.3} & 83.5 & 84.0 & 85.9 & \textbf{88.1} & 74.5 & 75.1 & 75.0 & \textbf{78.4} & 78.8 & 79.3 & 80.7 & \textbf{83.0}\\
\hline						
		\end{tabular}
					}
\end{table*}

We evaluate the influence of the resampling and reconstruction methods on the classification accuracy. Therefore, we determine the number of correctly predicted samples as \(N_T\) and the number of falsely predicted samples as \(N_F\). The sum of \(N_T+N_F\) yields the overall number of samples. The classification accuracy \(ACC\) is determined as the ratio of correctly predicted samples to the overall number of samples, thus
\begin{equation}
ACC = \frac{N_T}{N_T+N_F}.
\end{equation} 
As presented in the previous sections, we incorporated bilinear and bicubic interpolation as well as FSR for the state-of-the-art testing pipeline. In addition, we determined classification accuracies for our proposed pipeline incorporating the high-quality FSMR method. We evaluated the performance of the five presented networks for the widely used \textit{Caltech101} and \textit{Oxford-IIIT-Pet} datasets. Furthermore, the three presented loss patterns \textit{BLOCK, LINE, RAND} and the combination of all of them \textit{ALL} are incorporated. \\
In Table~\ref{tab:results_caltech} the results for the \textit{Caltech101} dataset are given. The table denotes the incorporated loss pattern in the first column. The second column determines the incorporated resampling technique in the training phase. The first row gives the considered network and the row below the used reconstruction and resampling methods in the testing phase. As we consider the case of having an already trained network available, where the incorporated interpolation technique in the training phase is unknown, we are interested in the best possible result for each combination of training interpolation and neural network. Thus, the best result per row and network is highlighted in bold face. The overall best for a network is additionally underlined. Considering EfficientNetB0, the \textit{Caltech101} dataset shows best classification accuracies if FSR is considered in the testing phase and bilinear or bicubic in the training phase. However, the overall best for three out of four loss patterns for EfficientNetB0 are achieved if our proposed pipeline with FSMR is used. For the larger networks of DenseNets and ResNets, our proposed training pipeline with the use of FSMR leads to highest classification accuracies in nearly all cases. The networks containing a higher number of parameters seem to profit more from a high-quality resampling in the preprocessing. In addition, one can observe that FSMR clearly  outperforms all other test interpolations. Assuming DenseNet121 and a \textit{LINE} loss pattern in the testing pipeline, FSMR improves the FSR result by \(11.0\)~percentage points if trained on lin. If trained on cub, the improvement is still \(7.0\)~percentage points and if already trained on FSMR, the gain is even larger with \(19.1\)~percentage points compared to the second-best performing test interpolation. The differences are huge for the large loss patterns \textit{BLOCK} and \textit{LINE} as considerable distortions are induced into the test images. The \textit{RAND} loss pattern shows a more unstructured loss pattern and thus, usually more neighborhood information is available for every reconstructed pixel. Thus, the gap between the incorporated test interpolations shrinks even though FSMR still remains the best performing test interpolation in nearly all cases. Furthermore, the overall best results for one network are also achieved with FSMR in the test phase and ideally if FSMR was already used for training. The networks profit from high-quality images already in the training phase.  The last row gives the average for each training interpolation. Once again, the best performing method is given in bold face. On average, FSMR is the best performing method for all networks. FSMR achieves on average a gain in classification accuracy of up to \(6.7\) percentage points for DenseNet121 compared to FSR. \\
The results for the \textit{Oxford-IIIT-Pet} dataset are given in Table~\ref{tab:results_pet}. The structure of the table is the same as in Table~\ref{tab:results_caltech}. The results for the four classes of loss patterns and all combinations of networks, training and test interpolation are shown. It can be observed that the results are not as consistent as the ones for the \textit{Caltech101} dataset. Nevertheless, if the last row in the Table~\ref{tab:results_pet} is examined, one can observe that FSMR is still the best performing method for all networks except EfficientNetB0. Thus, our proposed approach with a joint loss reconstruction and resampling is the preferred choice if high classification accuracies are desired. FSMR achieves an improvement of up to 3.3 percentage points on average for ResNet50.
\vspace{-.1cm}
\section{Conclusion}
In this paper, we could demonstrate that the usage of different interpolation methods for both, training and testing of a neural network has a high influence on the overall classification accuracy of networks. This is especially valid if losses occur during transmission that have to be reconstructed before classifiying the images and if the images have to be resized to one expected input size of a network. We highly recommend to conduct the reconstruction of losses and the resizing in one step jointly incorporating FSMR for the resampling before classification. Ideally, the networks were trained using FSMR for resizing and data augmentation techniques. Training on FSMR and incorporating FSMR for interpolation in the test phase yields best results for nearly all possible combinations of networks, datasets, and loss patterns. Independent of the training interpolation, FSMR yields on average best classification accuracies with gains of up to 6.7~percentage points. 
\vspace{-.2cm}
\section*{Acknowledgment}
This work was partly funded by the Deutsche Forschungsgemeinschaft (DFG, German Research Foundation) – SFB 1483 – \mbox{Project-ID~442419336}, EmpkinS.

\vspace{-.1cm}
\bibliographystyle{IEEEtran}
\bibliography{bib_4ivmsp2022.bib}

\end{document}

%% file: figures/DatabaseNew.tex
\begin{tikzpicture}
\tkzInit[xmax=4, ymax=4]
\begin{scope}[dashed]
\tkzGrid
\end{scope}
\draw[fill=black](0,0)circle(2pt);
\draw[fill=black](0,1)circle(2pt);
\draw[fill=black](0,2)circle(2pt);
\draw[fill=black](0,3)circle(2pt);	
\draw[fill=black](0,4)circle(2pt);	
\draw[fill=black](1,0)circle(2pt);
\draw[fill=black](1,1)circle(2pt);	
\draw[fill=black](1,2)circle(2pt);
\draw[fill=black](1,3)circle(2pt);	
\draw[fill=black](1,4)circle(2pt);
\draw[fill=black](2,0)circle(2pt);
\draw[fill=black](2,1)circle(2pt);
\draw[fill=black](2,2)circle(2pt);
\draw[fill=black](2,3)circle(2pt);	
\draw[fill=black](2,4)circle(2pt);	
\draw[fill=black](3,0)circle(2pt);
\draw[fill=black](3,1)circle(2pt);	
\draw[fill=black](3,2)circle(2pt);
\draw[fill=black](3,3)circle(2pt);	
\draw[fill=black](3,4)circle(2pt);
\draw[fill=black](4,0)circle(2pt);
\draw[fill=black](4,1)circle(2pt);	
\draw[fill=black](4,2)circle(2pt);
\draw[fill=black](4,3)circle(2pt);	
\draw[fill=black](4,4)circle(2pt);	

\draw[color=red](0.4, 0)circle(2pt);
\draw[color=red](0.8, 0)circle(2pt);
\draw[color=red](1.2, 0)circle(2pt);
\draw[color=red](1.6, 0)circle(2pt);
\draw[color=red](2.0, 0)circle(2pt);
\draw[color=red](2.4, 0)circle(2pt);
\draw[color=red](2.8, 0)circle(2pt);
\draw[color=red](3.2, 0)circle(2pt);
\draw[color=red](3.6, 0)circle(2pt);

\draw[color=red](0.4, 0.8)circle(2pt);
\draw[color=red](0.8, 0.8)circle(2pt);
\draw[color=red](1.2, 0.8)circle(2pt);
\draw[color=red](1.6, 0.8)circle(2pt);
\draw[color=red](2.0, 0.8)circle(2pt);
\draw[color=red](2.4, 0.8)circle(2pt);
\draw[color=red](2.8, 0.8)circle(2pt);
\draw[color=red](3.2, 0.8)circle(2pt);
\draw[color=red](3.6, 0.8)circle(2pt);

\draw[color=red](0.4, 1.6)circle(2pt);
\draw[color=red](0.8, 1.6)circle(2pt);
\draw[color=red](1.2, 1.6)circle(2pt);
\draw[color=red](1.6, 1.6)circle(2pt);
\draw[color=red](2.0, 1.6)circle(2pt);
\draw[color=red](2.4, 1.6)circle(2pt);
\draw[color=red](2.8, 1.6)circle(2pt);
\draw[color=red](3.2, 1.6)circle(2pt);
\draw[color=red](3.6, 1.6)circle(2pt);

\draw[color=red](0.4, 2.4)circle(2pt);
\draw[color=red](0.8, 2.4)circle(2pt);
\draw[color=red](1.2, 2.4)circle(2pt);
\draw[color=red](1.6, 2.4)circle(2pt);
\draw[color=red](2.0, 2.4)circle(2pt);
\draw[color=red](2.4, 2.4)circle(2pt);
\draw[color=red](2.8, 2.4)circle(2pt);
\draw[color=red](3.2, 2.4)circle(2pt);
\draw[color=red](3.6, 2.4)circle(2pt);

\draw[color=red](0.4, 3.2)circle(2pt);
\draw[color=red](0.8, 3.2)circle(2pt);
\draw[color=red](1.2, 3.2)circle(2pt);
\draw[color=red](1.6, 3.2)circle(2pt);
\draw[color=red](2.0, 3.2)circle(2pt);
\draw[color=red](2.4, 3.2)circle(2pt);
\draw[color=red](2.8, 3.2)circle(2pt);
\draw[color=red](3.2, 3.2)circle(2pt);
\draw[color=red](3.6, 3.2)circle(2pt);

\draw[color=red](0.4, 4)circle(2pt);
\draw[color=red](0.8, 4)circle(2pt);
\draw[color=red](1.2, 4)circle(2pt);
\draw[color=red](1.6, 4)circle(2pt);
\draw[color=red](2.0, 4)circle(2pt);
\draw[color=red](2.4, 4)circle(2pt);
\draw[color=red](2.8, 4)circle(2pt);
\draw[color=red](3.2, 4)circle(2pt);
\draw[color=red](3.6, 4)circle(2pt);



\end{tikzpicture}

%% file: figures/grid_with_hole.tex
\makeatletter
\tikzset{
    block filldraw/.style={
        draw, fill=black!10},
    block rect/.style={
        block filldraw, rectangle},
    block/.style={
        block rect, minimum height=0.8cm, minimum width=6em},
    from/.style args={#1 to #2}{
        above right={0cm of #1},
        /utils/exec=\pgfpointdiff
            {\tikz@scan@one@point\pgfutil@firstofone(#1)\relax}
            {\tikz@scan@one@point\pgfutil@firstofone(#2)\relax},
        minimum width/.expanded=\the\pgf@x,
        minimum height/.expanded=\the\pgf@y}}

\begin{tikzpicture}
\tkzInit[xmax=4, ymax=4]
\begin{scope}
\node[block rect, from={.8,.8 to 3.2,3.2}]{};
\end{scope}

\begin{scope}[dashed]
\tkzGrid
\end{scope}

\draw[fill=black](0,0)circle(2pt);
\draw[fill=black](0,1)circle(2pt);
\draw[fill=black](0,2)circle(2pt);
\draw[fill=black](0,3)circle(2pt);	
\draw[fill=black](0,4)circle(2pt);	
\draw[fill=black](1,0)circle(2pt);
\draw[fill=black](1,1)circle(2pt);	
\draw[fill=black](1,2)circle(2pt);
\draw[fill=black](1,3)circle(2pt);	
\draw[fill=black](1,4)circle(2pt);
\draw[fill=black](2,0)circle(2pt);
\draw[fill=black](2,1)circle(2pt);
\draw[fill=black](2,2)circle(2pt);
\draw[fill=black](2,3)circle(2pt);	
\draw[fill=black](2,4)circle(2pt);	
\draw[fill=black](3,0)circle(2pt);
\draw[fill=black](3,1)circle(2pt);	
\draw[fill=black](3,2)circle(2pt);
\draw[fill=black](3,3)circle(2pt);	
\draw[fill=black](3,4)circle(2pt);
\draw[fill=black](4,0)circle(2pt);
\draw[fill=black](4,1)circle(2pt);	
\draw[fill=black](4,2)circle(2pt);
\draw[fill=black](4,3)circle(2pt);	
\draw[fill=black](4,4)circle(2pt);

\draw[color=red](0,0)circle(3pt);
\draw[color=red](0,1)circle(3pt);
\draw[color=red](0,2)circle(3pt);
\draw[color=red](0,3)circle(3pt);	
\draw[color=red](0,4)circle(3pt);	
\draw[color=red](1,0)circle(3pt);
\draw[color=red](1,4)circle(3pt);
\draw[color=red](2,0)circle(3pt);
\draw[color=red](2,4)circle(3pt);	
\draw[color=red](3,0)circle(3pt);
\draw[color=red](3,4)circle(3pt);
\draw[color=red](4,0)circle(3pt);
\draw[color=red](4,1)circle(3pt);	
\draw[color=red](4,2)circle(3pt);
\draw[color=red](4,3)circle(3pt);	
\draw[color=red](4,4)circle(3pt);


\end{tikzpicture}

%% file: figures/flow_graph.tex
\tikzstyle{block} = [rectangle, draw, fill=white!15, 
    text width = \columnwidth, text centered, rounded corners, anchor = south west] 
\tikzstyle{block1} = [rectangle, draw, fill=white!15, 
    text width = .18\columnwidth, text centered, rounded corners, anchor = south east] 
\tikzstyle{block11} = [rectangle, draw=red, fill=white!15, 
    text width = .4\columnwidth, text centered, rounded corners, anchor = south east] 
\tikzstyle{block12} = [rectangle, draw, fill=white!15, 
    text width = .4\columnwidth, text centered, rounded corners, anchor = south east] 
\tikzstyle{block21} = [rectangle, draw=blue, fill=white!15, 
    text width = .45\columnwidth, text centered, rounded corners, anchor = south west] 
\tikzstyle{block22} = [rectangle, draw, fill=white!15, 
    text width = .45\columnwidth, text centered, rounded corners, anchor = south east, dashed] 
        
\begin{tikzpicture}

\node[block21] at (0,6.5) (Train) {Training Set};
\node[block22] at (\columnwidth,6.5) (Val) {Validation Set};

\node[block] at (0,5) (Interp) {Data Augmentation\\ Rotation \& Zoom};
\node[block] at (0,4) (int1) {Resample};
\node[block] at (0,3) (res1) {Resize};
\node[block] at (0,2) (int2) {Resample};

%

\node[block] at (0,0.5) (Train_net) {Training of network};

\draw [blue, ->] (Train.south) -- (Train|-Interp.north);
\draw [blue, ->] (Train.south|-Interp.south) -- (Train|-int1.north);
\draw [blue, ->] (Train.south|-int1.south) -- (Train|-res1.north);
\draw [blue, ->] (Train.south|-res1.south) -- (Train|-int2.north);
\draw [blue, ->] (Train.south|-int2.south) -- (Train|-Train_net.north);

\draw [dashed, ->] (Val.south) -- (Val|-Interp.north);
\draw [dashed, ->] (Val.south|-Interp.south) -- (Val|-int1.north);
\draw [dashed, ->] (Val.south|-int1.south) -- (Val|-res1.north);
\draw [dashed, ->] (Val.south|-res1.south) -- (Val|-int2.north);
\draw [dashed, ->] (Val.south|-int2.south) -- (Val|-Train_net.north);

%
%

\end{tikzpicture}

%% file: figures/test_setup_old.tex
\tikzstyle{block} = [rectangle, draw, fill=white!15, 
    text width = .45\columnwidth, text centered, rounded corners, anchor = south west] 
\tikzstyle{block1} = [rectangle, draw, fill=white!15, 
    text width = \columnwidth, text centered, rounded corners, anchor = south east] 
\tikzstyle{block11} = [rectangle, draw=red, fill=white!15, 
    text width = \columnwidth, text centered, rounded corners, anchor = south east] 
\tikzstyle{block12} = [rectangle, draw, fill=white!15, 
    text width = \columnwidth, text centered, rounded corners, anchor = south east] 
\tikzstyle{block21} = [rectangle, draw=blue, fill=white!15, 
    text width = .2\columnwidth, text centered, rounded corners, anchor = south west] 
\tikzstyle{block22} = [rectangle, draw, fill=white!15, 
    text width = .2\columnwidth, text centered, rounded corners, anchor = south west, dashed] 
        
\begin{tikzpicture}

\node[block11] at (.5\columnwidth,6.5) (Test) {Test Set};


\node[block1] at (.5\columnwidth,5) (res2) {Add pattern};
\node[block1] at (.5\columnwidth,4) (int3) {Reconstruct};
\node[block1] at (.5\columnwidth,3) (patt) {Resize};
\node[block1] at (.5\columnwidth,2) (recon) {Resample};
%

\node[block12] at (.5\columnwidth,0.5) (Test_net) {Testing of network};

%
%

\draw [red,->] (res2.north|-Test.south) --(res2); 
\draw [red,->] (res2) -- (int3) node[pos=0.5, right, color=black, circle, draw, inner sep=0.5pt]{\small1};
\draw [red,->] (int3) -- (patt) node[pos=0.5, right, color=black, circle, draw, inner sep=0.5pt]{\small2};
\draw [red,->] (patt) -- (recon);
\draw [red,->] (recon) -- (recon.south|-Test_net.north) node[pos=0.5, right, color=black, circle, draw, inner sep=0.5pt]{\small3};

%

\end{tikzpicture}

%% file: figures/test_setup_proposed.tex
\tikzstyle{block} = [rectangle, draw, fill=white!15, 
    text width = .45\columnwidth, text centered, rounded corners, anchor = south west] 
\tikzstyle{block1} = [rectangle, draw, fill=white!15, 
    text width = \columnwidth, text centered, rounded corners, anchor = south east] 
\tikzstyle{block11} = [rectangle, draw=red, fill=white!15, 
    text width = \columnwidth, text centered, rounded corners, anchor = south east] 
\tikzstyle{block12} = [rectangle, draw, fill=white!15, 
    text width = \columnwidth, text centered, rounded corners, anchor = south east] 
\tikzstyle{block21} = [rectangle, draw=blue, fill=white!15, 
    text width = .2\columnwidth, text centered, rounded corners, anchor = south west] 
\tikzstyle{block22} = [rectangle, draw, fill=white!15, 
    text width = .2\columnwidth, text centered, rounded corners, anchor = south west, dashed] 
        
\begin{tikzpicture}

\node[block11] at (.5\columnwidth,6.5) (Test) {Test Set};
%

%
\node[block1] at (.5\columnwidth,5) (resfsmr) {Add pattern};
\node[block1] at (.5\columnwidth,3) (pattfsmr) {Resize};
\node[block1] at (.5\columnwidth,2) (reconfsmr) {Reconstruct and Resample};

\node[block12] at (.5\columnwidth,0.5) (Test_net) {Testing of network};

%
%

%
\draw [red,->] (resfsmr.north|-Test.south) --(resfsmr);
\draw [red,->] (resfsmr) -- (pattfsmr) node[pos=0.2, right, color=black, circle, draw, inner sep=0.5pt]{\small1};
\draw [red,->] (pattfsmr) -- (reconfsmr) ;
\draw [red,->] (reconfsmr) -- (reconfsmr.south|-Test_net.north) node[pos=0.5, right, color=black, circle, draw, inner sep=0.5pt]{\small3*};
%

\end{tikzpicture}

%% file: figures/grid_with_hole_resized.tex
\makeatletter
\tikzset{
    block filldraw/.style={
        draw, fill=black!10},
    block rect/.style={
        block filldraw, rectangle},
    block/.style={
        block rect, minimum height=0.8cm, minimum width=6em},
    from/.style args={#1 to #2}{
        above right={0cm of #1},
        /utils/exec=\pgfpointdiff
            {\tikz@scan@one@point\pgfutil@firstofone(#1)\relax}
            {\tikz@scan@one@point\pgfutil@firstofone(#2)\relax},
        minimum width/.expanded=\the\pgf@x,
        minimum height/.expanded=\the\pgf@y}}

\begin{tikzpicture}
\tkzInit[xmax=4, ymax=4]
\begin{scope}
\node[block rect, from={1.2,1.5 to 2.8,2.5}]{};
\end{scope}

\begin{scope}[dashed]
\tkzGrid
\end{scope}

\draw[fill=black](0,0)circle(2pt);
\draw[fill=black](0,1)circle(2pt);
\draw[fill=black](0,2)circle(2pt);
\draw[fill=black](0,3)circle(2pt);	
\draw[fill=black](0,4)circle(2pt);	
\draw[fill=black](1,0)circle(2pt);
\draw[fill=black](1,1)circle(2pt);	
\draw[fill=black](1,2)circle(2pt);
\draw[fill=black](1,3)circle(2pt);	
\draw[fill=black](1,4)circle(2pt);
\draw[fill=black](2,0)circle(2pt);
\draw[fill=black](2,1)circle(2pt);
\draw[fill=black](2,2)circle(2pt);
\draw[fill=black](2,3)circle(2pt);	
\draw[fill=black](2,4)circle(2pt);	
\draw[fill=black](3,0)circle(2pt);
\draw[fill=black](3,1)circle(2pt);	
\draw[fill=black](3,2)circle(2pt);
\draw[fill=black](3,3)circle(2pt);	
\draw[fill=black](3,4)circle(2pt);
\draw[fill=black](4,0)circle(2pt);
\draw[fill=black](4,1)circle(2pt);	
\draw[fill=black](4,2)circle(2pt);
\draw[fill=black](4,3)circle(2pt);	
\draw[fill=black](4,4)circle(2pt);

\draw[color=red](0.4, 0)circle(2pt);
\draw[color=red](0.8, 0)circle(2pt);
\draw[color=red](1.2, 0)circle(2pt);
\draw[color=red](1.6, 0)circle(2pt);
\draw[color=red](2.0, 0)circle(2pt);
\draw[color=red](2.4, 0)circle(2pt);
\draw[color=red](2.8, 0)circle(2pt);
\draw[color=red](3.2, 0)circle(2pt);
\draw[color=red](3.6, 0)circle(2pt);

\draw[color=red](0.4, 0.8)circle(2pt);
\draw[color=red](0.8, 0.8)circle(2pt);
\draw[color=red](1.2, 0.8)circle(2pt);
\draw[color=red](1.6, 0.8)circle(2pt);
\draw[color=red](2.0, 0.8)circle(2pt);
\draw[color=red](2.4, 0.8)circle(2pt);
\draw[color=red](2.8, 0.8)circle(2pt);
\draw[color=red](3.2, 0.8)circle(2pt);
\draw[color=red](3.6, 0.8)circle(2pt);

\draw[color=red](0.4, 1.6)circle(2pt);
\draw[color=red](0.8, 1.6)circle(2pt);
\draw[color=red](3.2, 1.6)circle(2pt);
\draw[color=red](3.6, 1.6)circle(2pt);

\draw[color=red](0.4, 2.4)circle(2pt);
\draw[color=red](0.8, 2.4)circle(2pt);
\draw[color=red](3.2, 2.4)circle(2pt);
\draw[color=red](3.6, 2.4)circle(2pt);

\draw[color=red](0.4, 3.2)circle(2pt);
\draw[color=red](0.8, 3.2)circle(2pt);
\draw[color=red](1.2, 3.2)circle(2pt);
\draw[color=red](1.6, 3.2)circle(2pt);
\draw[color=red](2.0, 3.2)circle(2pt);
\draw[color=red](2.4, 3.2)circle(2pt);
\draw[color=red](2.8, 3.2)circle(2pt);
\draw[color=red](3.2, 3.2)circle(2pt);
\draw[color=red](3.6, 3.2)circle(2pt);

\draw[color=red](0.4, 4)circle(2pt);
\draw[color=red](0.8, 4)circle(2pt);
\draw[color=red](1.2, 4)circle(2pt);
\draw[color=red](1.6, 4)circle(2pt);
\draw[color=red](2.0, 4)circle(2pt);
\draw[color=red](2.4, 4)circle(2pt);
\draw[color=red](2.8, 4)circle(2pt);
\draw[color=red](3.2, 4)circle(2pt);
\draw[color=red](3.6, 4)circle(2pt);

\end{tikzpicture}